\documentclass[10pt, prd, twocolumn]{revtex4}

\usepackage{amsmath}
\usepackage{amssymb}
\usepackage{graphicx}
\usepackage{bm} 

\begin{document}

\title{Black hole thermodynamics 
with the cosmological constant as independent variable:
Bridge between the enthalpy and the Euclidean path
integral approaches}

\author{Jos\'{e} P. S. Lemos}
\affiliation{Centro de de Astrof\'{\i}sica e Gravita\c{c}\~{a}o -
CENTRA, Departamento de
F\'{\i}sica, Instituto Superior T\'{e}cnico - IST,
Universidade de Lisboa -
UL, Avenida Rovisco Pais 1, 1049-001, Portugal}
\email{joselemos@ist.utl.pt}

\author{Oleg B. Zaslavskii}
\affiliation{Department of Physics and Technology, Kharkov V. N.
Karazin National
University, 4 Svoboda Square, Kharkov 61022, Ukraine, and Institute of
Mathematics and Mechanics, Kazan Federal University, 18 Kremlyovskaya St.,
Kazan 420008, Russia}
\email{zaslav@ukr.net}

\begin{abstract}

Viewing the cosmological constant $\Lambda<0$ as an independent
variable, we consider the thermodynamics of the Schwarzschild black
hole in an anti-de Sitter (AdS) background.  For this system, there is
one approach which regards the enthalpy as the master thermodynamic
variable and makes sense if one considers the vacuum pressure due to
the cosmological constant acting in the volume inside the horizon and
the outer size of the system is not restricted.  From this approach a
first law of thermodynamics emerges naturally.  There is yet another
approach based on the Euclidean action principle and its path integral
that puts the black hole inside a cavity, defines a quasilocal energy
at the cavity's boundary, and from which a first law of thermodynamics
in a different version also emerges naturally. The first approach has
affinities with critical phenomena in condensed matter physics and the
second approach is an ingredient necessary for the construction of
quantum gravity. The bridge between the two approaches is carried out
rigorously, putting thus the enthalpic thermodynamics
with $\Lambda$ as independent variable on the same footing as the
quasilocal energy approach.

\end{abstract}

\maketitle

\section{Introduction}

In recent years, a new direction in gravitational and black hole
thermodynamics appeared.  It is based on treatment of the cosmological
constant $\Lambda <0$ as a thermodynamic variable and leads to a
number of interesting physical consequences in which the thermodynamic
potential enthalpy emerges naturally and a close analogy with van der
Waals forces and critical phenomena in condensed matter physics can be
carried out. The works and reviews on the subject can be found in
\cite{krt,dolan0,cgkp,km,dol,wugeliu,k}.  These nontrivial features
arise in cases like the Schwarzschild and Reissner-Nordstr\"{o}m black
holes in an anti-de Sitter (AdS) background \cite{km}. A main feature
is the first law of thermodynamics in which the term $d\Lambda$ is
taken into account with $\Lambda$ acting as a vacuum pressure
$P$. Indeed $P$ takes the value $P=-\frac{\Lambda}{8\pi}$ and is
conjugate to a thermodynamic volume $V$ given by $V=\frac{4}{3}\pi
r_{+}^{3}$, $r_+$ being the horizon radius.  Substantiation of this
approach is based on precise derivations as well as heuristic
arguments that take into account the volume inaccessible to an
observer due to the existence of a horizon, whereas the outer size of
the black hole thermodynamic system is not restricted in this
approach.

On the other hand, there exists a well-defined gravitational black
hole thermodynamics based on the Euclidean action principle and path
integral approach which is recognized
as an ingredient necessary for the construction of
a quantum gravity \cite{hartleh,bolt,hawkpage}. This approach was put
on a firm basis taking into account the finiteness of the system as an
essential ingredient, i.e., the black hole
is constrained to lie inside a
cavity. At the cavity's boundary a temperature $T$ is prescribed,
defining thus a canonical ensemble
\cite{york,martyork,can,brownyork,mann}, or if electric charge is
present a grand canonical ensemble should be used \cite{ch,jp}.  Also
at the cavity's boundary one can define a quasilocal energy and from
it extract the first law of thermodynamics.  In this approach the
horizon turns into a bolt \cite{bolt} and the volume under the horizon
becomes totally irrelevant, whereas the outer size of the black hole
thermodynamic system is required to have a well defined boundary.

One approach uses the enthaply as the master thermodynamic variable,
makes explicit the importance of the volume inside of the black hole
horizon, and the outer size is not constrained. The other approach uses
the quasilocal energy as the master thermodynamic variable, cuts the
inside of the black hole horizon as unimportant, and puts a boundary in
the outer region. Clearly, the two approaches look dual to each
other. Thus, the existence of these  two dual independent thermodynamic
formulations with different concepts for the same system
deserves a careful study and comparison.

The aim of the present work
is to make the bridge between these two formulations and thus put the
thermodynamics with $\Lambda$ as an independent variable
with the enthalpy as the master thermodynamic variable approach
on the same
footing as the quasilocal energy canonical ensemble approach.
We consider a  Schwarzschild-AdS
black hole and give a rigorous
derivation of the corresponding first law of thermodynamics
in enthalpic terms 
from the first law of the finite size gravitational thermodynamics. In
doing so, we substantiate, why and how the thermodynamic volume
$\frac{4}{3} \pi r_{+}^{3}$ appears.

\section{The spacetime}

Let us consider the Schwarzschild-AdS spacetime black hole
with metric
\begin{equation}
ds^{2}=-f(r)dt^{2}+\frac{dr^{2}}{f(r)}+r^{2}(d\theta ^{2}+\sin ^{2}\theta d\phi
^{2}),
\end{equation}
with
\begin{equation}
f(r)=1-\frac{2M}{r}-\frac{\Lambda r^{2}}{3}\,,
\label{metricf1}
\end{equation}
where $M$ is the ADM mass and $\Lambda$ is
the cosmological constant. The
cosmological constant is negative,
$\Lambda <0$ so that spacetime is
asymptotically AdS.
Equation $f(r)=0$ has one root $r_{+}$ that corresponds to the event
horizon, i.e., $ 1-\frac{2M}{r_+}-\frac{\Lambda r_+^{2}}{3}=0$, and so
the mass $M$ is given in terms of $r_+$ and $\Lambda$ as
\begin{equation}
M=\frac{r_{+}}{2}\left(1-\frac{\Lambda r_{+}^{2}}{3}\right)\,.
\label{mmass}
\end{equation}
The metric function (\ref{metricf1}) in terms of $r_+$ is then
\begin{equation}
f(r)=1-\frac{r_{+}}{r}+\frac{\Lambda }{3}
\left(\frac{r_{+}^{3}}{r}-r^{2}\right)\,.
\label{r+}
\end{equation}

\section{Thermodynamics in a cavity in the Euclidean action
quasilocal energy approach}

Let us consider a black hole in the canonical ensemble, so
the temperature $T$ is
fixed on a sphere
with area $A$ and radius $R$, with
$A=4\pi R^{2}$. The corresponding
thermodynamics formalism
was initiated
in \cite{york} for a Schwarzschild black hole in a cavity
(see also \cite{martyork,can}),
extended in \cite{brownyork} where
quasilocal energy was defined, and applied for
an AdS
black hole in a cavity
\cite{mann} (for the electric charge
extensions see \cite{ch,jp}). 
In this approach one uses
the quasilocal energy $E$ as the master thermodynamic
variable.

The quasilocal
energy at a sphere with radius $R$,
can be envisaged as a function of the mass $M$, the radius $R$,
and $\Lambda$, i.e.,  
$E=E(M,R,\Lambda)$. We are thus treating the cosmological
constant as an independent variable that can change
its value thermodynamically.
Following \cite{brownyork,mann} we can write $E=E(M,R,\Lambda)$ as
\begin{equation}
E(M,R,\Lambda)=E_{0}(R,\Lambda)-R\sqrt{f(M,R,\Lambda)}\,, \label{e2}
\end{equation}
where from Eq.~(\ref{metricf1})
\begin{equation}
f(M,R,\Lambda)=1-\frac{2M}{R}-\frac{\Lambda R^{2}}{3}\,.  \label{red2}
\end{equation}
and $E_{0}(R,\Lambda)$ is an appropriate reference spacetime
with respect to which the
energy is calculated. Its meaning consists in subtracting
from the energy $E(M,R,\Lambda)$ of the relevant spacetime
the energy $E_{0}(R,\Lambda)$
of a
reference spacetime with the same boundary data.

The quasilocal
energy at a sphere with radius $R$,
can be envisaged  also as a function of the radius $r_+$, the radius $R$,
and $\Lambda$, i.e.,  
$E(r_+,R,\Lambda)$,
and written as 
\begin{equation}
E(r_+,R,\Lambda)=E_{0}(R,\Lambda )-R\sqrt{f(r_+,R,\Lambda)}\,,  \label{e1}
\end{equation}
where  from Eq.~(\ref{r+})
\begin{equation}
f(r_+,R,\Lambda)=1-\frac{r_{+}}{R}+\frac{\Lambda }{3}
\left(\frac{r_{+}^{3}}{R}-R^{2}\right)\,,  \label{red1}
\end{equation}
and again $E_{0}(R,\Lambda)$ is an appropriate reference spacetime
with respect to which the
energy is calculated.

By construction
\begin{equation}
f(r_+,R,\Lambda)=f(M,R,\Lambda)\equiv f(R)\,.  \label{red12}
\end{equation}
So in what follows we use $f(R)$ to simplify the notation.

The Bekenstein-Hawking entropy calculated
in this formalism is 
\cite{york,martyork,can,brownyork,mann,ch,jp} 
\begin{equation}
S=\pi r_{+}^{2}\,.
\label{entropy}
\end{equation}
So, clearly, from Eq.~(\ref{entropy}) we see that
in Eq.~(\ref{e1})
we can exchange $r_+$ for $S$ straightforwardly.
This exchange is useful in some situations.

\section{First law of thermodynamics I:
$E(M,R,\Lambda)$
(or $E(M,A,\Lambda)$)}

Seeing the quasilocal energy as $E(M,R,\Lambda)$,
or equivalently as $E(M,A,\Lambda)$
since $A=4\pi R^2$,
allows us to  write formally the first law
of thermodynamics as 
\begin{equation}
dE=\mu dM
-pdA-l d\Lambda\,,
\label{f1}
\end{equation}
where $\mu$, $p$, and $l$ are defined by
\begin{equation}
\mu=\left(\frac{\partial E}{\partial M}\right)_{A,\Lambda}
 \,,
\label{f11}
\end{equation}
\begin{equation}
p=
-\left( \frac{\partial E}{\partial A}\right)_{M,\Lambda}  \,,
\label{f12}
\end{equation}
\begin{equation}
l=-\left(\frac{\partial E}{\partial \Lambda }\right)_{M,A}
 \,.
\label{f13}
\end{equation}
The quantity $\mu$ is the thermodynamic variable
conjugate to $M$ at the sphere of radius $R$, $p$ is the surface pressure
at the sphere of radius $R$, and
$l$ is the thermodynamic variable
conjugate to $\Lambda$ at the sphere of radius $R$, it has units of volume.
Explicitly, from Eqs.~(\ref{e2})-(\ref{red2})
and  Eqs.~(\ref{f11})-(\ref{f13}) 
we have
\begin{equation}
\mu=
\frac{1}{\sqrt{f(R)}}\,,
\label{mu}
\end{equation}
\begin{equation}
8\pi Rp=\frac{1-\frac{M}{R}-
\frac{2\Lambda R^2}{3}}
{\sqrt{f(R)}}
-\left( \frac{\partial E_{0}}{\partial R}
\right)_{\Lambda}\text{,}\label{p2}
\end{equation}
\begin{equation}
l=
-\frac{R^{3}}{6\sqrt{f(R)}}-\left( \frac{
\partial E_{0}}{\partial \Lambda }\right) _{R}\,.
\label{lfinal}
\end{equation}
All three quantities depend on the redshift factor
$f(R)$, and $p$ and $l$ depend on $E_{0}$.

\section{First law of thermodynamics II:
$E(r_+,R,\Lambda)$
(or $E(S,A,\Lambda)$)}

Seeing the quasilocal energy as $E(r_+,R,\Lambda)$,
or equivalently as $E(S,A,\Lambda)$
since $S=4\pi r_+^2$ from Eq.~(\ref{entropy})
and $A=4\pi R^2$,
we can write formally the first law
of thermodynamics as 
\begin{equation}
dE=TdS-pdA-\lambda d\Lambda\,,  \label{f0}
\end{equation}
where 
\begin{equation}
T =\left( \frac{\partial E}{\partial S }\right)_{A,\Lambda}\,,
\label{f001}
\end{equation}
\begin{equation}
p =-\left( \frac{\partial E}{\partial A }\right)_{r_+,\Lambda}\,,
\label{f002}
\end{equation}
\begin{equation}
\lambda =-\left( \frac{\partial E}{\partial \Lambda }\right)_{r_+,A}\,.
\label{f003}
\end{equation}
$T$ is the temperature
at the sphere of radius $R$, $p$ is the surface pressure
at the sphere of radius $R$, 
and $\lambda$ is the thermodynamic quantity
conjugate to $\Lambda$ at the sphere of radius $R$, with units of volume.
Explicitly, using $S=4\pi r_+^2$, see Eq.~(\ref{entropy}),
we have  from Eq.~(\ref{f001}) that
$T =\left( \frac{\partial E}{\partial S }\right)_{A,\Lambda}=\frac{1}{2\pi r_+}
\left( \frac{\partial E}{\partial r_+ }\right)_{R,\Lambda}$.
Then, using  Eqs.~(\ref{e1})-(\ref{red1}),
we find
\begin{equation}
T= \frac{
{\cal T}
}{\sqrt{f(R)}}\,,\,\quad\,\,
{\cal T}= \frac{1-\Lambda r_+^2}{4\pi r_+}\,.
\label{x}
\end{equation}
Using $A=4\pi R^2$, 
we have that Eq.~(\ref{f002}) can be written as
$p=-\left( \frac{\partial E}{\partial A }\right)_{r_+,\Lambda}=
-\frac{1}{8\pi R}\left( \frac{\partial E}{\partial R}
\right)_{r_+,\Lambda}$, and so from Eqs.~(\ref{e1})-(\ref{red1}),
we find
\begin{equation}
8\pi Rp=\frac{1-\frac{r_{+}}{2R}+\frac{\Lambda
}{3}\left(\frac{r_{+}^{3}}{2R}-2R^{2}\right)}
{\sqrt{f(R)}}-\left( \frac{\partial E_{0}}{\partial R}
\right)_{\Lambda}\text{,}\label{p1}
\end{equation}
Finally, from Eq.~(\ref{f003}) and using  Eqs.~(\ref{e1})-(\ref{red1}),
we find
\begin{equation}
\lambda =\frac{r_{+}^{3}-R^{3}}{6\sqrt{f(R)}}-\left( \frac{\partial
E_{0}}{ \partial \Lambda }\right)_{R}\,.
\label{lambdafinal}
\end{equation}
All three quantities depend on the redshift factor
$f(R)$. In particular Eq.~(\ref{x}) means that
the temperature $T$ at $R$ is given by the Tolman formula,
i.e., it is given by some temperature $\cal T$
redshifted to $R$, where $\cal T$ can be considered
the Hawking temperature.
Note that $p$ and $\lambda$ depend on $E_{0}$.

\section{First law of thermodynamics III: Enthalpy $H(S,P)$}

We are now in a position to find a thermodynamics description
with the enthalpy as the master thermodynamic variable.

For that we equate Eqs.~(\ref{f1}) and (\ref{f0}).
But first note that Eqs.~(\ref{p2}) and (\ref{p1}) 
give the same surface pressure $p$
as it is seen from
Eq.~(\ref{mmass}), i.e.,
$
1-\frac{M}{R}-
\frac{2\Lambda R^2}{3}=
1-\frac{r_{+}}{2R}+\frac{\Lambda
}{3}\left(\frac{r_{+}^{3}}{2R}-2R^{2}\right)$.
Then
equating Eqs.~(\ref{f1}) and (\ref{f0})
the two terms $p\,dA$ cancel out
and 
one finds
\begin{equation}
\mu \,dM
=TdS+
\left(l-\lambda
\right)d\Lambda\,.  \label{f1first}
\end{equation}
Eq.~(\ref{mu}) gives $\mu=
\frac{1}{\sqrt{f(R)}}$, and from Eq.~(\ref{x})
one gets $T=
\frac{\cal T}{\sqrt{f(R)}}$.
Also, from Eqs.~(\ref{lfinal}) and (\ref{lambdafinal}), one has
$l-\lambda=
-\left[\left(\frac{\partial E}{\partial \Lambda }\right)_{M,A}
-\left( \frac{\partial E}{\partial \Lambda }\right)_{r_+,A}
\right]
=-\frac{r_+^3}{6\sqrt{f(R)}}
$. The dependence on $E_0$ has now disappeared.
Collecting these expressions into 
Eq.~(\ref{f1first})
one finds,
\begin{equation}
dM={\cal T}dS+VdP\,,  \label{1}
\end{equation}
where
\begin{equation}
V=\frac{4}{3}\pi r_{+}^{3}\,,
\end{equation}
and 
\begin{equation}
P=-\frac{\Lambda }{8\pi }\,.
\end{equation}
The quantity $V$ formally coincides with the volume of a region in a flat
spacetime inside a radius $r_{+}$, and usually it is
referred to as the thermodynamic volume. The quantity 
$P$ is positive since $\Lambda$ is negative, and 
has the meaning of 
vacuum pressure. It is
clear from Eq.~(\ref{1}) that the mass $M$ can be identified with the enthalpy
$H$, 
$M=H$ and so, 
\begin{equation}
dH={\cal T}dS+VdP\,,  \label{1final}
\end{equation}
so that $H=H(S,P)$. 
Thus,  as we have just showed,
the enthalpy approach
is directly connected to
the
quasilocal energy path
integral Euclidean approach and vice-versa.

\section{Conclusions}

We showed that the first law of black hole thermodynamics in the form
of Eq.~(\ref{1}) with $\Lambda$ as a thermodynamic variable and the
enthalpy $H$ as the master thermodynamic variable does indeed follow
from the black hole gravitational finite-size thermodynamics based on
the quasilocal energy Euclidean action principle. In this sense, in
making this bridge, we substantiated the approach discussed in
\cite{krt,dolan0,cgkp,km,dol,wugeliu,k}. We explained the
identification of the mass $M$ with the enthalpy $H$.  We clarified
why the first law of thermodynamics, Eq.~(\ref{1}), does not depend on
the reference point, i.e., the term $E_{0}$ disappears.  As it follows
from the process of derivation, we have not appealed to what is hidden
under the horizon. Nonetheless, we demonstrated why and how a
thermodynamic volume $V$ 
emerges consistently in the first
law with the expression for this volume being
$V=\frac{4}{3}\pi r_{+}^{3}$.  It is also worth noting that the
original set of variables included the surface pressure $p$ related to
the boundary and we traced how $p$ is replaced by the completely
different bulk vacuum pressure $P$.

It would be interesting to generalize the present results to the
electric charge and rotating cases.

\begin{acknowledgments}

JPSL
thanks  Funda\c c\~ao para a Ci\^encia e Tecnologia (FCT), Portugal,
for financial support through Grant~No.~UID/FIS/00099/2013
and Grant No.~SFRH/BSAB/128455/2017, and Coordena\c
c\~ao de Aperfei\c coamento do Pessoal de N\'\i vel Superior (CAPES),
Brazil, for support within the Programa CSF-PVE, Grant
No.~88887.068694/2014-00.  OBZ thanks support from SFFR, Ukraine,
Project No.~32367 and thanks also Kazan Federal University for a state
grant for scientific activities.

\end{acknowledgments}

\end{document}